\documentclass[pra,aps,reprint,groupedaddress,showpacs]{revtex4-1}
\usepackage{graphicx}
\usepackage{amsmath}
\usepackage{amsfonts}
\usepackage{amssymb}
\usepackage{amsthm}
\usepackage{color}

\begin{document}
\title{Dissipation in quantum turbulence in superfluid $^4$He above 1K}

\author{J. Gao}
\affiliation{National High Magnetic Field Laboratory, 1800 East Paul Dirac Drive, Tallahassee, FL 32310, USA}
\affiliation{Mechanical Engineering Department, Florida State University, Tallahassee, FL 32310, USA}

\author{W. Guo}
\affiliation{National High Magnetic Field Laboratory, 1800 East Paul Dirac Drive, Tallahassee, FL 32310, USA}
\affiliation{Mechanical Engineering Department, Florida State University, Tallahassee, FL 32310, USA}

\author{S. Yui}
\affiliation{Department of Physics, Osaka City University, 3-3-138 Sugimoto, Sumiyoshi-Ku, Osaka 558-8585, Japan}

\author{M. Tsubota}
\affiliation{Department of Physics, Osaka City University, 3-3-138 Sugimoto, Sumiyoshi-Ku, Osaka 558-8585, Japan}

\author{W.F. Vinen}
\affiliation{School of Physics and Astronomy, University of Birmingham, Birmingham B15 2TT, United Kingdom}

\date{\today}

\begin{abstract}
There are two commonly discussed forms of quantum turbulence in superfluid $^4$He above 1K: in one there is a random tangle of quantizes vortex lines, existing in the presence of a non-turbulent normal fluid;  in the second there is a coupled turbulent motion of the two fluids,  often exhibiting quasi-classical characteristics on scales larger than the separation between the quantized vortex lines in the superfluid component.  The decay of vortex line density, $L$, in the former case is often described by the equation $dL/dt=-\chi_2 (\kappa/2\pi)L^2$, where $\kappa$ is the quantum of circulation,  and $\chi_2$ is a dimensionless parameter of order unity.  The decay of total turbulent energy, $E$, in the second case is often characterized by an effective kinematic viscosity, $\nu'$, such that $dE/dt=-\nu' \kappa^2 L^2$. We present new values of $\chi_2$ derived from numerical simulations and from experiment,  which we compare with those derived from a theory developed by Vinen and Niemela. We summarise what is presently known about the values of $\nu'$ from experiment, and we present a brief introductory discussion of the relationship between $\chi_2$ and $\nu'$,  leaving a more detailed discussion to a later paper.
\end{abstract}

\maketitle

\section{Introduction}
Below about 2.17 K, liquid $^4$He becomes a superfluid,  in which an inviscid irrotational superfluid component coexists with a viscous normal-fluid component \cite{tilley_book}. Any vorticity in the superfluid component is confined to quantized vortex lines,  each of which carries a single quantum of circulation $\kappa=h/m$, where $h$ is Planck's constant and $m$ is the mass of a He atom \cite{donnelly_book}. Flow in each of the two fluids can be turbulent. Turbulence in the superfluid component (quantum turbulence) takes the form of an irregular tangle of vortex lines which interact with each other and,  through a force of ``mutual friction",   with the normal fluid \cite{vinen_niemela2002a}. Turbulence in the normal fluid is similar to that in a classical fluid,  but modified by the mutual friction.  Dissipation,  associated with viscosity,  plays an important role in classical turbulence;  notably, in providing a sink where the energy flux in a high Reynolds number Richardson cascade can be absorbed at small length scales.  It must play a similarly important role in quantum turbulence,  although,  as we shall see,  dissipative mechanisms are then more complex than in the classical case.

Except at temperatures well below 1K,  where the normal fluid has disappeared, dissipation in the turbulent superfluid component is due,  as we shall see, to the mutual friction. If we ignore a small transverse (non-dissipative) component,  the force of mutual friction per unit length of vortex line can be expressed in terms of a dimensionless parameter $\alpha$ \cite{vinen_niemela2002a}.  Except at temperatures very close to the superfluid transition temperature, $\alpha$ is significantly less than unity, with the result that vortex line motion is determined largely by vortex-vortex interactions,  the mutual friction leading to only a relatively slow shrinkage in the total length, $L$,  of vortex line per unit volume.  Dissipation in the normal fluid is due to both mutual friction and viscosity.

It is the aim of this paper to discuss these forms of dissipation for two commonly studied types of quantum turbulence (QT),  the dissipation being observed in the free decay of the turbulence.

QT can be most easily produced by a heat current,  which is carried in superfluid helium by a counterflow of the two fluids,  and this is the form of QT that was first subject to detailed experimental study \cite{vinen1957d,vinen1957c,vinen1957a}.  It was thought for many years that this \textit{thermal counterflow turbulence (TCT)} involved only the superfluid component,  and took the form of a more or less random vortex tangle, for which the turbulent energy is confined to scales comparable with or less than the average spacing, $\ell=L^{-1/2}$,  between the vortex lines. The corresponding energy spectrum, $E_Q(k)$, has the form
\begin{equation} E_Q(k)=\frac{\rho_s\kappa^2}{4\pi\rho \ell^2 k}f\Big(\frac{k\ell}{2\pi}\Big),
\label{Eq1}
\end{equation}
where the function $f(x)$ depends on the precise form of the ``random tangle", but tends to unity for large $x$,  and tends rapidly to zero for $x<1$ \cite{nemirovskii2013a}.  $\rho_s/\rho$ is the superfluid fraction.  It was suggested, on dimensional and physical grounds \cite{vinen1957a}, that, when the heat current is switched off, the line density might decay as
\begin{equation} \frac{dL}{dt}=-\frac{\chi_2 \kappa}{2\pi}L^2,
\label{Eq2}
\end{equation}
where $\chi_2$ is a dimensionless parameter of order unity. Noting that the energy per unit mass associated with a random tangle of vortex lines is given by
\begin{equation}   E_Q=\int_0^\infty E_Q(k)dk \approx \frac{\rho_s \kappa^2}{4\pi \rho}L\ln{\frac{\ell}{\xi_0}},
\label{Eq3}
\end{equation}
where $\xi_0$ is the vortex core parameter, we see that the turbulent energy per unit mass would then decay as
\begin{equation} \frac{dE_r}{dt}=-\nu'_v \kappa^2 L^2, \quad  \frac{\nu'_v}{\kappa}=\frac{\chi_2 \rho_s}{8\pi^2 \rho}\ln{\frac{\ell}{\xi_0}},
\label{Eq4}
\end{equation}
where $\nu'_v$ is an effective kinematic viscosity.

Recent experiments \cite{marakov2015a,jetpletter}, based on the use of He$_2^{*}$ excimer molecules as tracers of the normal-fluid flow, have shown that this form of QT, involving only what we shall call a \textit{random vortex tangle}, exists in TCT only at sufficiently small heat fluxes; at larger heat fluxes the tangle is accompanied by turbulence in both fluids on scales up to the size of the containing channel. We shall write the resulting energy spectrum as
\begin{equation}
E(k)=E_Q(k) + E_{Cs}(k) + E_{Cn}(k),
\label{Eq5}
\end{equation}
where $E_Q(k)$ is still given by Eq.(\ref{Eq1}), $E_{Cs}(k)$ is produced by partial polarization of the vortex lines,  and $E_{Cn}(k)$ relates to the turbulent energy in the normal fluid. In the steady state this large-scale turbulence in the two fluids is partially coupled and has an energy spectrum, $E(k)\propto k^{-n}$ on scales significantly larger than $\ell$,  where the exponent $n$ varies with the heat flux but is always larger that the Kolmogorov value \cite{kolmogorov1941a}, $5/3$ \cite{marakov2015a};  that $n>5/3$ is a reflection of the fact that coupling is incomplete,  so that there is dissipation on all length scales \cite{jetpletter}. After the source of heat is turned off,  the heat flux in the channel decays to zero in a time given by  a thermal RC time constant (typically 10 ms).  Then the two fluids become fully coupled in a similar time, retaining for a time the $k^{-n}$ energy on large scales. Finally, over a further period of typically 1-10 s, the energy spectrum on large scales evolves into the form expected for in a classical inertial-range Richardson cascade;  i.e. a Kolmogorov spectrum, $E(k)\propto k^{-5/3}$ \cite{kolmogorov1941a}.

We emphasize three points relating to the fully coupled turbulence: as long as full coupling is maintained, there is no dissipation due to mutual friction;  the large scale non-dissipative motion in the superfluid component is generated by a partial polarization of the vortex lines;  and large scale motion in the normal  component  is non-dissipative because the viscosity of the normal fluid is sufficiently small. As we shall see more clearly later, dissipation can occur in both fluids on scales comparable with or less than $\ell$, that in the superfluid component  being due to mutual friction, partial decoupling having occurred,  and that in the normal component being due  a combination of viscosity and mutual friction. Because dissipation on scales of order $\ell$ is now much more complicated than is the case if the turbulence is confined to the superfluid component and to scales of order $\ell$, Eq.(\ref{Eq4}) need no longer apply.

The decay of line density associated with large-scale coupled turbulence was first studied by Stalp \textit{et al} \cite{Stalp1999a}, the coupled turbulence having been generated in the wake of a moving grid.  These authors showed that their experimental results could be explained in purely classical terms,  if it was assumed that there was at all times a Richardson-Kolmogorov cascade ($E(k)\propto k^{-5/3}$), terminated at small scales by dissipation described by the equation
\begin{equation}  \frac{dE_C}{dt}=-\epsilon=-\nu'\kappa^2 L^2,
\label{Eq6}
\end{equation}
 where $\nu'$ is another effective kinematic viscosity;  $E_C$ is the total quasi-classical turbulent energy,  given by integrating $E_{Cs}(k) + E_{Cn}(k)$ over $k$ (the contribution of $E_Q(k)$ to the total energy is small and can be neglected). Stalp \textit{et al} argued that Eq.(\ref{Eq6}) is the analogue of the expression $\nu\langle \omega^2 \rangle$ for dissipation in classical homogeneous turbulence,  where $\langle \omega^2 \rangle$ is the mean square classical vorticity. We emphasize that, although the expressions (\ref{Eq4}) and (\ref{Eq6}) for the rate of decay of turbulent energy are similar in form,  they relate to different physical situations, and in neither case has there been any really rigorous discussion of their validity.  Furthermore,  as we shall discuss later, the two effective kinematic viscosities,  $\nu'_v$ and $\nu'$ need not have the same value.  In future we shall refer to large-scale coupled turbulence of the type produced by flow through grid,  or in the decay of strongly excited TCT at large times, as \textit{quasi-classical quantum turbulence}.

 We remark here that a Kolmogorov energy spectrum can,   strictly speaking,  apply only to a steady state in which energy is fed in continuously at some large scale $D$ at a rate $\epsilon$;  there is then a constant energy flux, equal to $\epsilon$, down an inertial sub-range,  $2\pi/D\gg k \gg 2\pi/\ell$,  within which the energy spectrum has the full Kolmogorov form $E(k)\sim \epsilon^{2/3} k^{-5/3}$ (we are ignoring the effects of intermittency) . In decaying turbulence the energy flux,  $\epsilon$, cannot be strictly independent of either time or wave number,  so that the Kolmogorov dependence on wave number, $k^{-5/3}$, cannot be strictly correct.  In practice,  however,  most of the energy is often concentrated in the largest eddies (wave numbers close to $2\pi/D$),  so that $\epsilon$ is independent of $k$,  for $k>2\pi/D$, to a reasonable approximation;  and the decay is sufficiently slow that the Kolmogorov spectrum holds with a slowly decreasing value of $\epsilon$.

 Except perhaps for a simple theoretical calculation of $\chi_2$,  reviewed later in Section III, there has so far been hardly any serious theoretical justification for the two forms of dissipation,  and for many years even experimental justification was inadequate.  Similarly it has proved difficult to derive reliable values of the two effective kinematic viscosities from experiment.  In the case of $\nu'_v$ (or equivalently $\chi_2$) there had been no careful study of the decay of TCT at heat currents sufficiently small that there was no large-scale turbulence.  In the other case values of $\nu'$ were obtained from observations of the decay of vortex line density combined with questionable assumptions about the form of the large-scale energy spectrum as it relates to turbulence in a channel of finite cross-section.  Only very recently has $\nu'$ been determined in a more satisfactory way for the case of decaying TCT \cite{Gao2016a},  although the results have yet to be compared carefully with those obtained solely from the decay of vortex line density.  The general aims of this paper are, as far as possible, to remedy these various shortcomings.

 The results of our new experiments on the decay of a random vortex tangle and our measurements of $\chi_2$  are described in Section II. In Section III we summarize an existing theory of $\chi_2$,  assess its likely validity, and compare its predictions with experiment. In Section IV we describe the numerical simulations relating to a random vortex tangle,  and we compare the results with the experiment and with the theory of Section III.   In Section V we present a critical summary of our present knowledge of the experimental values of the effective kinematic viscosity $\nu'$, and in Section VI we present a brief introductory theoretical discussion of the relationship between $\chi_2$ (or $\nu'_v$) and $\nu'$,  leaving a more serious discussion of what is actually a difficult problem to a later paper.  We present an overall summary of our work in Section VII.

\section{Dissipation in a random vortex tangle: the experimental measurement of $\chi_2$.}
Our new experiments on the decay of vortex line density associated with TCT have been based on the observed attenuation of second sound, using what is now a standard technique,  as described in,  for example, references \cite{vinen1957c,babuin2015a}.  The actual apparatus is identical with that described in reference \cite{marakov2015a}.

\begin{figure}[ht]
\begin{center}
\includegraphics[scale=0.6]{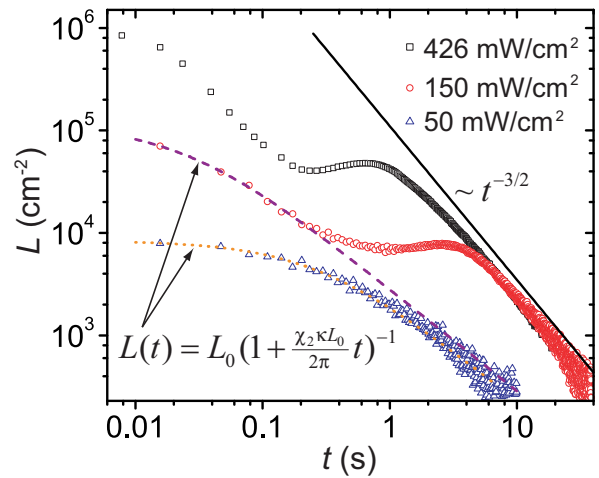}
\caption{(color online) Observed decays of vortex line density in decaying TCT (1.65K)}.
\label{Fig1}
\end{center}
\end{figure}

As we have explained,  the form of decay of vortex line density given by Eq.(\ref{Eq2}) can be expected to be observed in the decay of TCT only if the steady heat flux is small enough to ensure that there is no large-scale turbulence.  This is indeed the case is evident in the decay shown by the lowest line in Fig.\ref{Fig1}.

\begin{figure}[ht]
\begin{center}
\includegraphics[scale=0.62]{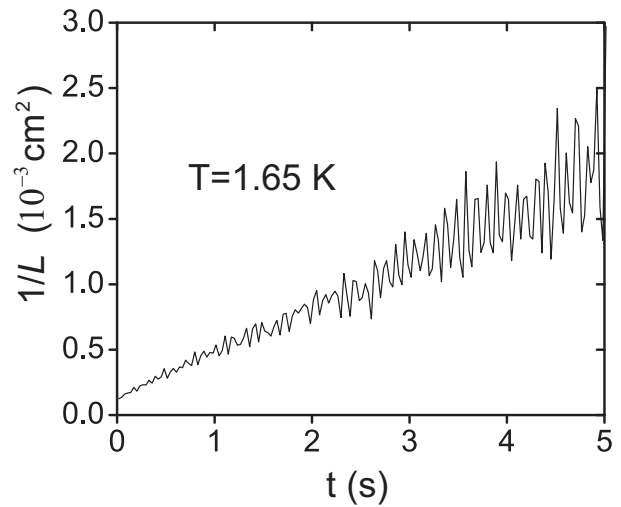}
\caption{Observed decay in line density from a small heat flux.}
\label{Fig2}
\end{center}
\end{figure}

In Fig.\ref{Fig2} we show data for a decay from a small heat flux plotted in a form,  $(1/L)$ \textit{versus} $t$, which serves to demonstrate more clearly that Eq.(\ref{Eq2}) is indeed obeyed.

\begin{figure}[ht]
\begin{center}
\includegraphics[scale=0.6]{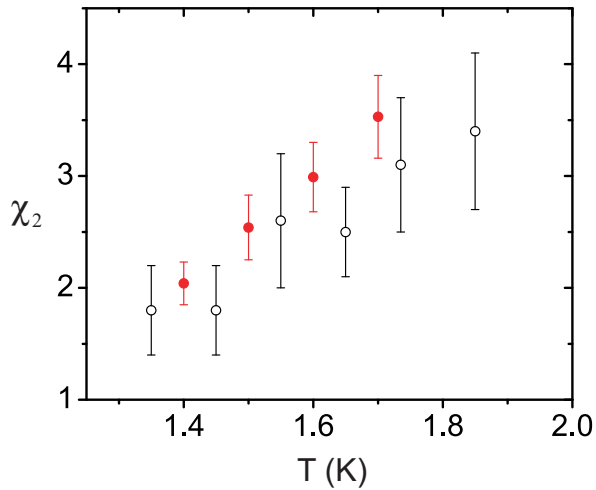}
\caption{(color online) Observed (open circles) and theoretical (filled circles) values of $\chi_2$,  the theoretical values being derived from Eq.(\ref{Eq12}).}
\label{Fig3}
\end{center}
\end{figure}

Values of $\chi_2$ deduced from decays of this type are shown as a function of temperature in Fig.\ref{Fig3}.

\section{Dissipation in a random vortex tangle:  a theory of $\chi_2$}
In this section we shall summarise a theory of $\chi_2$ that was proposed by Vinen and Niemela \cite{vinen_niemela2002a},  and we shall compare the results with experiment.

We assume that the force of mutual friction per unit length of vortex line is given by
\begin{equation}   \mathbf{f}=-\rho_s \kappa \alpha \hat{\kappa} \times [\hat{\kappa} \times (\mathbf{v}_n - \mathbf{v}_L)],
\label{Eq7}
\end{equation}
where $\hat{\kappa}$ is a unit vector along the length of the vortex and $\mathbf{v}_L$ is the velocity with which the vortex moves perpendicular to its length.  We have neglected any transverse component of the mutual friction. We shall further assume that during the decay described by Eq.(\ref{Eq2}) the normal fluid is at rest, apart from the local dragging by a moving\^{a}\^{a} vortex that is incorporated into the definition of the mutual friction parameter $\alpha$ \cite{vinen_hall2}. Dissipation is then due entirely to mutual friction. Finally,  we shall assume that the magnitude of $\mathbf{v}_L$ is given to a good enough approximation by the local induction approximation
\begin{equation}
v_L = \frac{\kappa}{4\pi R} \ln{\Big(\frac{R}{\xi_0}\Big)},
\label{Eq8}
\end{equation}
where $R$ is the local radius of curvature of the vortex,  and $\xi_0$ is the vortex core parameter. In other words, we have neglected the effect of both long-range interactions and the force of mutual friction itself on the motion of a vortex.  It follows that the rate of dissipation of energy per unit mass of helium  is given by
\begin{equation}  \frac{dE_r}{dt} = -\frac{\rho_s}{\rho} \kappa \alpha L \langle v_L^2 \rangle =-\alpha\Big(\frac{\rho_s \kappa^3}{16\pi^2 \rho}\Big) \langle\Big[\frac{1}{R^2}\Big(\ln{\frac{R}{\xi_0}}\Big)^2\Big]\rangle L,
\label{Eq9}
\end{equation}
where $\langle ... \rangle$ denotes an average over the vortex tangle. We neglect the slow variation of the logarithmic term with $L$,  putting $R=R_0\approx\ell$ in that term,  and we follow Schwarz \cite{schwarz1988a} by assuming that
\begin{equation}  \langle\Big[\frac{1}{R^2}\Big]\rangle =c_2^2 L,
\label{Eq10}
\end{equation}
where $c_2$ depends only on temperature.  It follows that
\begin{equation}   \frac{\nu'_v}{\kappa} = \frac{\alpha c_2^2 \rho_s}{16\pi^2 \rho}\Big[\ln{\frac{\ell}{\xi_0}}\Big]^2,
\label{Eq11}
\end{equation}
and therefore
\begin{equation}  \chi_2 = \frac{\alpha c_2^2}{2} \ln{\frac{\ell}{\xi_0}}.
\label{Eq12}
\end{equation}

This derivation is based on three assumptions:  that,  as we have mentioned,  there is no motion of the normal fluid; that the vortex lines form a random tangle; and that use of the local induction approximation is justified.  We shall present an argument in favour of the first assumption in Section VI.  The other assumptions seem reasonable.

Values of $\chi_2$ derived from Eq.(\ref{Eq12}) are included in Fig.\ref{Fig3}.  The required values of $c_2$ are taken from the simulations of the steady state described in Section IV,  and values of $\alpha$ are taken from reference \cite{d-b-tables}.  We see that within the error bars there is agreement with experiment.

\section{Dissipation in a random vortex tangle:  simulations relating to $\chi_2$}

A brief report of simulations leading a verification of the form of the decay of line density and to values of $\chi_2$ at a temperature of 1.9 K has already been published \cite{mineda2013c}.  Here we present the results of a more detailed studies,  covering a range of temperatures,  first for the case of spatially uniform flows,  and then for flows between solid boundaries.

\subsection{The steady state for spatially uniform flows.}

For a given temperature we must first simulate the steady state counterflow, for two reaons. It is from these states that the decays must start,  and we can determine whether values of the parameter $c_2$,  obtained for the steady state, lead \textit{via} Eq.(\ref{Eq12}) to agreement with experimentally observed values of $\chi_2$.

Our numerical simulation is based on the vortex filament model with the full Biot-Savart integral. We carry out simulations for spatially uniform flows  in a cubical box,  side 1 mm,  with periodic boundary conditions in all directions.   We replace the vortex lines by a discrete set of points with minimum spatial resolution $\Delta \xi = 8.0 \times 10^{-4}$ cm.  We integrate in time with a fourth-order Runge-Kutta scheme with time resolution $\Delta t = 1.0 \times 10^{-4}$ s.  The initial state is a set of randomly oriented vortex loops of radius $0.23$ mm.  The spatially uniform applied velocities satisfy the condition of no net mass flow $\rho_n v_n + \rho_s v_{s,a} = 0$.  We have checked that any contribution to the net superflow from the evolving vortex tangle is negligible in comparison with $v_{s,a}$.  The parameters used in the simulations are shown in Table~\ref{Table-1}.

\begin{table}[htb]
\caption{Parameters used in numerical simulations.}
\label{Table-1}
\begin{tabular}{|p{1.5cm}|p{1.5cm}|p{1.5cm}|p{1.5cm}|}
\hline
T & $\alpha$ & $\alpha'$ & $v_n$ \\
K &  & & mm s$^{-1}$ \\
\hline
1.4 & 0.052 & 0.017 & 9.0 \\
1.5 & 0.073 & 0.018 & 7.0 \\
1.6 & 0.098 & 0.016 & 6.0 \\
1.7 & 0.127 & 0.012 & 5.0 \\
\hline
\end{tabular}
\end{table}

\begin{figure}[ht]
\begin{center}
\includegraphics[scale=0.8]{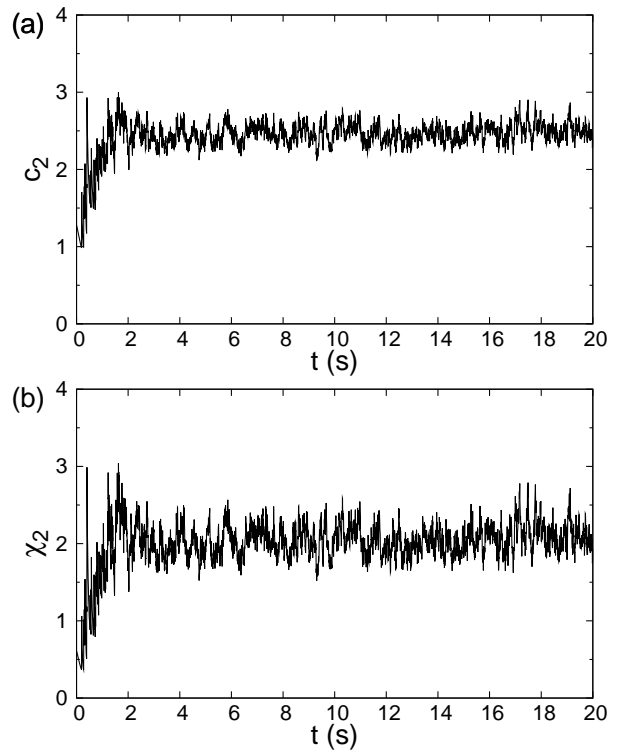}
\caption{Value of $c_2$ derived from simulations of the approach of counterflow to a steady state, and the corresponding value of $\chi_2$ derived from Eq.(\ref{Eq12}).}
\label{Fig4}
\end{center}
\end{figure}

We run the simulations for 20 s.  The vortex line density, $L$, is found to reach a steady average value, $L_0$,  with fluctuations,  in about 5 s. The parameter $c_2$,  calculated from the Eq.(\ref{Eq10}) and the equation

\begin{equation}  \langle \frac{1}{R^2} \rangle = \frac{1}{\Omega L} \int \frac{d \xi}{R^2},
\label{Eq13}
\end{equation}
where $\Omega$ is the volume of the numerical box, together with the values of $\chi_2$ derived from Eq.(\ref{Eq12}), are shown as a function of time for a temperature of 1.4 K in Fig. \ref{Fig4}. As we see,  they too reach steady states after a few seconds, but with significant fluctuations.  The relatively large fluctuations have their origin in the relatively small computational box;  a  larger box would require prohibitively long computer runs.   We have performed similar simulations for several temperatures,  the results of which are summarized in terms of time-averages in Table~\ref{Table-2}. The computed values of $\chi_2$ in Fig. \ref{Fig3} were taken from Table~\ref{Table-2}.

We emphasize that the theoretical/computational values of $\chi_2$ plotted in Fig.\ref{Fig3} were derived from Table~\ref{Table-2};  the agreement with experiment was therefore evidence that Eq.(\ref{Eq12}) is at least approximately valid if the values of $c_2$ are taken from numerical simulations of the steady state.  We must now turn to numerical simulations of the decaying turbulence,  to check whether the simulated decays obey Eq.(\ref{Eq2}) with values of $\chi_2$ that agree with those in Table~\ref{Table-2}.

\begin{table}[htb]
\caption{Statistically steady values of the vortex line density, $L_0$, the parameter $c_2$, the mean radius of curvature, $R_0$,  and the corresponding values of $\chi_2$ derived from Eq.(\ref{Eq12}).}
\label{Table-2}
\begin{tabular}{|c|c|c|c|c|c|}
\hline
$T$ & $v_n$ & $L_0$ & $c_2$ & $R_0$ & $\chi_2$ \\
K & cm/s & $10^3$ cm$^{-2}$ & & $10^{-3}$ cm & \\
\hline
$1.4$ & $0.9$ & $6.54\pm 0.03$ & $2.47\pm 0.12$ & $5.03\pm 0.25$ & $2.04\pm 0.19$ \\
$1.5$ & $0.7$ & $5.80\pm 0.24$ & $2.31\pm 0.13$ & $5.71\pm 0.32$ & $2.54\pm 0.29$ \\
$1.6$ & $0.6$ & $6.14\pm 0.25$ & $2.16\pm 0.12$ & $5.93\pm 0.31$ & $2.99\pm 0.31$ \\
$1.7$ & $0.5$ & $6.34\pm 0.30$ & $2.06\pm 0.11$ & $6.12\pm 0.33$ & $3.53\pm 0.37$ \\
$1.4$ & $0.7$ & $3.59\pm 0.29$ & $2.47\pm 0.21$ & $6.82\pm 0.59$ & $2.10\pm 0.34$ \\
$1.4$ & $1.1$ & $10.0\pm 0.27$ & $2.44\pm 0.08$ & $4.10\pm 0.13$ & $1.97\pm 0.13$ \\
\hline
\end{tabular}
\end{table}

\subsection{Decays from spatially uniform flows}

In these simulations the applied velocities,  $v_n$ and $v_{s,a}$, are turned off at time $t=0$,  and the way in which the line density decays with $t$ is determined. Data are averaged over 30 decays at each temperature.

\begin{figure}[ht]
\begin{center}
\includegraphics[scale=0.6]{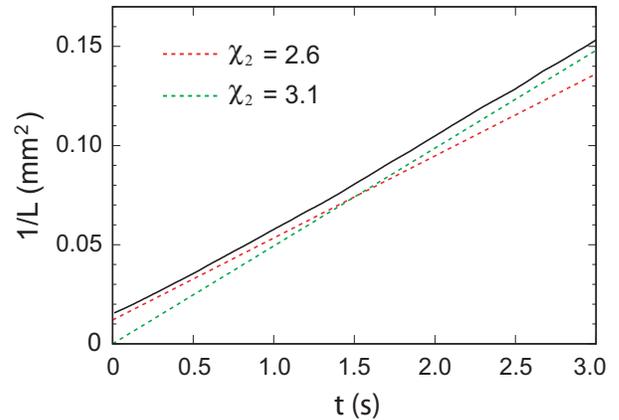}
\caption{(color online) ($1/L$) plotted against time from simulations at $T=1.4$K and $v_n=9$ mm s$^{-1}$.}
\label{Fig5}
\end{center}
\end{figure}

Fig.\ref{Fig5} shows the way in which the simulated line density decays with time at 1.4K,  in the form of a plot of $1/L$ against time.

\begin{figure}[ht]
\begin{center}
\includegraphics[scale=0.8]{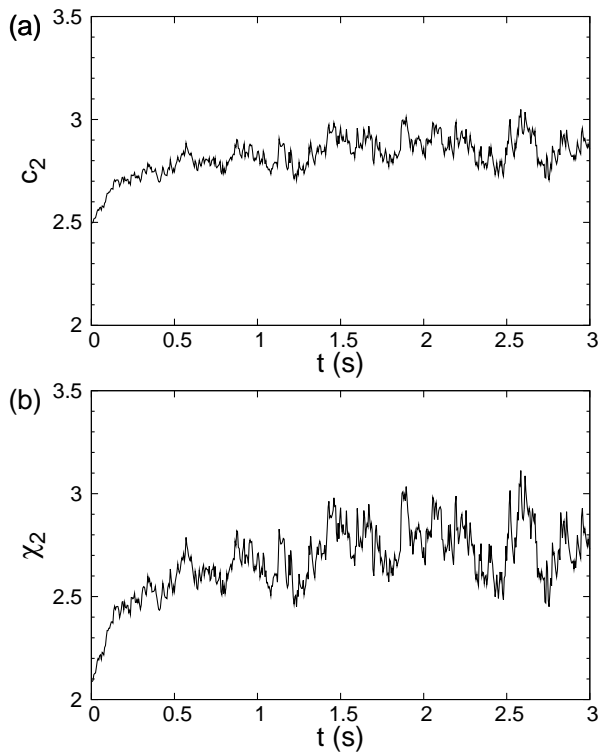}
\caption{a) The variation of the parameter $c_2$ with time from simulations of the decaying line density at $T=1.4$K and $v_n=9$ mm s$^{-1}$. (b) The variation of $\chi_2$ with time,  obtained by substituting $c_2$ from Fig.\ref{Fig6}a into Eq.(\ref{Eq12}).}
\label{Fig6}
\end{center}
\end{figure}

We see that,  in contrast to the corresponding experimental decay (Fig.\ref{Fig2}), Eq.(\ref{Eq2}) is apparently not obeyed;  the slope of the plotted line,  which ought to be proportional to the constant $\chi_2$ increases markedly with time (the values of $\chi_2$ are also too large). The increase at times greater than about 1 s may be due in part to the vortex line density becoming too small (the ratio of line spacing to the spatial period has become greater than about 0.3),  and in part to the effect of the logarithmic factor in Eq.(\ref{Eq12}).   A possible explanation of the discrepancy at smaller times is that the parameter $c_2$ in Eq.(\ref{Eq12}) changes during the simulated decay.  That $c_2$ does indeed change during the simulated decay is shown in Fig.\ref{Fig6}(a); furthermore, as we see from Fig.\ref{Fig6}(b),  this changing $c_2$ leads \textit{via} Eq.(\ref{Eq12}) to a changing value of $\chi_2$ that would lead,  at least qualitatively, to a decay curve with the shape shown in Fig.\ref{Fig5}. Similar results emerge from simulations at other temperatures.

\begin{figure}[ht]
\begin{center}
\includegraphics[scale=1.2]{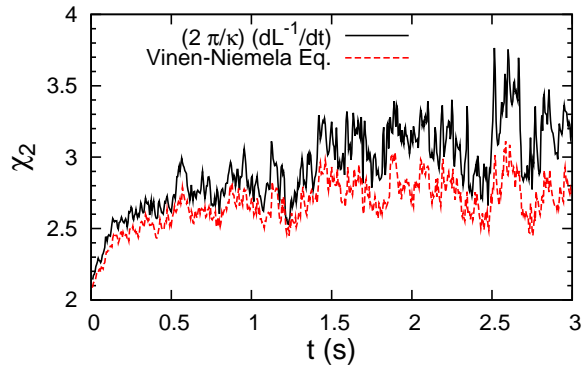}
\caption{(color online) Plots of $\chi_2$ against time derived as explained in the text.}
\label{Fig7}
\end{center}
\end{figure}

That the variation with time of the slope of the line in Fig.\ref{Fig5} is indeed due to the variation with time of the parameter $c_2$ is shown more strikingly in Fig.\ref{Fig7},  where we compare on the same graph the time dependence of the value $\chi_2$ derived both by differentiating $1/L$ in Fig.\ref{Fig5} with respect to time and by substituting the value of $c_2$ from Fig.\ref{Fig6}(a) into Eq.(\ref{Eq12}). Even the random fluctuations of $c_2$ are reflected to a significant degree in fluctuations in $\chi_2$ derived from Fig.\ref{Fig5}. The situation at other temperatures is similar.  We conclude then that the theory underlying Eq.(\ref{Eq12}) is in reasonably good agreement with the results of the simulations,  but not,  to a significant extent,  with experiment.  This suggests strongly that some factor relevant to the experiments is missing from both the theory and the simulations.  A possible candidate for this factor is the fact that,  in contrast to the theory and the simulations,  the experiments relate to flow in a channel of finite cross-section. We investigate this possibility in the next sub-section.

\subsection{Decays from flows in channels of finite cross-section}

Simulations relating to decaying counterflow in two types of channel have been carried out: one is formed between two parallel solid boundaries,  separated by 1 mm;  the other is a channel with square (1 mm $\times$ 1 mm) cross-section,  again with solid boundaries.  The conditions at the solid boundaries are that the normal fluid velocity vanishes,  and that the normal component of the superfluid velocity vanishes. Otherwise,  periodic boundary conditions are applied. In the steady state a parabolic velocity profile in the normal fluid is prescribed. Here we shall present only the results for two parallel boundaries;  the results for the channel with square cross section are broadly similar,  but are less clear cut because of large fluctuations in the line density in the steady state.
Data relating to the steady states in the case of the parallel plates are displayed in Table~\ref{Table-3}.

\begin{table}[htb]
\caption{Parameters analogous to those in Table~\ref{Table-2}, for flow between parallel plates.}
\label{Table-3}
\begin{tabular}{|c|c|c|c|c|c|}
\hline
$T$ & $v_n$ & $L_0$ & $c_2$ & $R_0$ & $\chi_2$ \\
K & cm s$^{-1}$ & $10^3$ cm$^{-2}$ & & $10^{-3}$ cm & \\
\hline
$1.4$ & $1.1$ & $5.94\pm 0.64$ & $2.17\pm 0.14$ & $6.03\pm 0.37$ & $1.60\pm 0.20$ \\
$1.5$ & $0.9$ & $5.67\pm 0.43$ & $2.02\pm 0.14$ & $6.61\pm 0.41$ & $1.97\pm 0.26$ \\
$1.6$ & $0.8$ & $6.58\pm 0.70$ & $1.84\pm 0.12$ & $6.76\pm 0.39$ & $2.19\pm 0.28$ \\
$1.7$ & $0.7$ & $6.74\pm 0.67$ & $1.79\pm 0.11$ & $6.86\pm 0.41$ & $2.68\pm 0.33$ \\
\hline
\end{tabular}
\end{table}

\begin{figure}[ht]
\begin{center}
\includegraphics[scale=0.9]{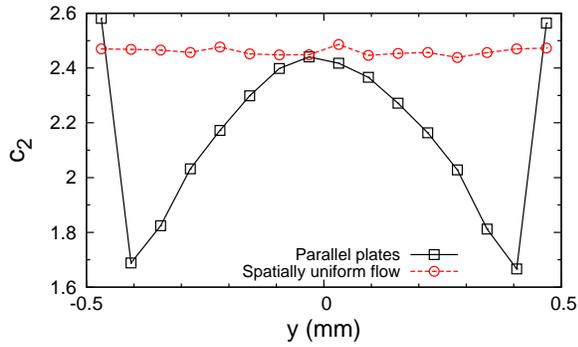}
\caption{(color online) Plots showing how $c_2$,  averaged over time in the steady state, varies with position across the channel. Blue line: flow between parallel plates;  red line:  spatially uniform flow. Temperature = 1.4 K.}
\label{Fig8}
\end{center}
\end{figure}

\begin{figure}[ht]
\begin{center}
\includegraphics[scale=0.75]{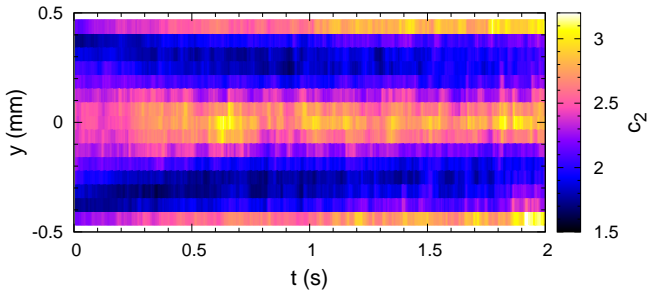}
\caption{(color online) Diagram showing how $c_2$ varies with position across the channel and with time during a decay. Temperature = 1.4 K.}
\label{Fig9}
\end{center}
\end{figure}

Before we proceed further we recall that the presence of solid boundaries in the steady state is known from simulations to lead to severe spatial inhomogeneity in the vortex line density \cite{Aarts-deWaele1994a, Baggaley2013a, Baggaley2014a, Yui-Tsubota2015a};  the vortex line density is greatly enhanced near the boundary (values of $L_0$,  and other parameters, in Table~\ref{Table-3} are spatial averages). We must therefore enquire whether there is also inhomogeneity in the value of $c_2$.  That there is indeed such inhomogeneity is shown in Figs \ref{Fig8} and \ref{Fig9},  derived from the simulations.  We see that the parameter $c_2$ is strongly reduced in regions where the vortex line density is increased,  and that this reduction persists in time during a decay.

\begin{figure}[ht]
\begin{center}
\includegraphics[scale=0.9]{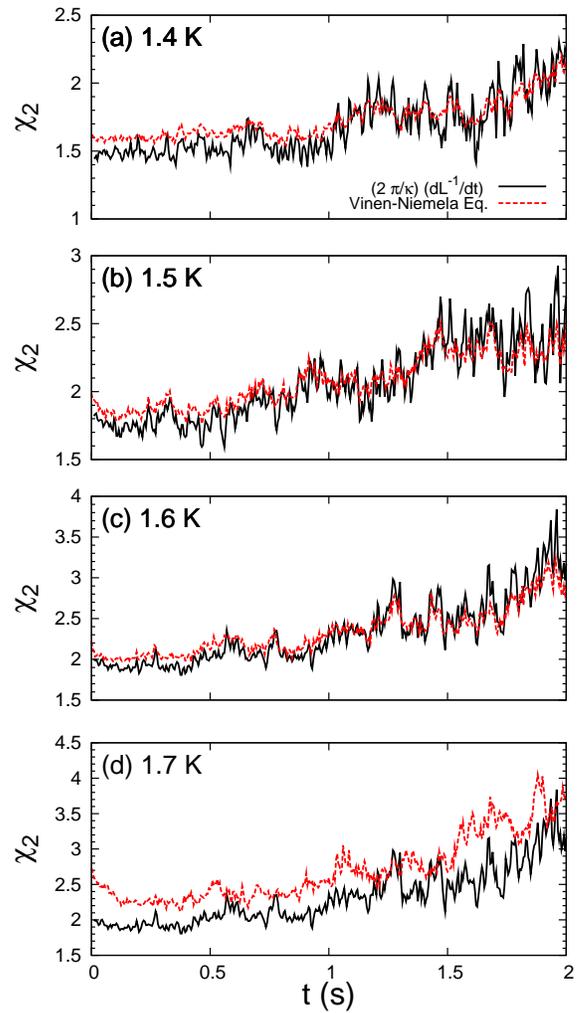}
\caption{(color online) Plots of $\chi_2$ against time for flow between parallel plates.}
\label{Fig10}
\end{center}
\end{figure}

The experimental observations of $\chi_2$,  reported in Section II,  relate to the decay of spatially averaged vortex line densities.  Our simulations of the decays between parallel plates lead to the corresponding values of $\chi_2$ that are displayed in Fig.\ref{Fig10},  where they are compared with the predictions of Eq.(\ref{Eq12}),  in which we have substituted spatially averaged values of the parameter $c_2$ taken from our simulations. We see that the agreement between the simulations and the predictions of Eq.(\ref{Eq12}) is still good and provides further confirmation that the theory of Section III is valid.  Furthermore, for times less than 1 s,  the variation with time of $\chi_2$ has largely disappeared,  and that the actual values of $\chi_2$ are in better agreement with experiment. This improved agreement with experiment is comforting and suggests that boundary effects are important in determining values of $c_2$ and therefore the precise form of the decays. However, reservations must be emphasized.  It is now clear that values of $c_2$ are quite sensitive to the precise form of the flows,  and our simulations still relate to flows that are not exactly the same as those in our experiments.  The experiments \cite{marakov2015a} use wider channels; in practice the velocity profile of the normal fluid differs generally from the Poiseuille form \cite{marakov2015a};  and in practice the vortex lines in the superfluid component are likely to suffer drag or pinning at the solid boundaries.  Simulations that take account of these differences are starting to be practicable (Yui, Tsubota and Kobayashi, to be published), and could eventually allow more satisfactory comparison with experiment.

\section{Dissipation in quasi-classical quantum turbulence:  experimental values of $\nu'$}
We turn now to the decay of large-scale coupled turbulence, as observed in the wake of flow through a grid and in the decay of TCT when the steady heat flux is large. We shall not be concerned with the early stages in these decays.  In the case of grid turbulence,  it has been supposed \cite{Stalp1999a} that a Kolmogorov spectrum is established quickly, with energy-containing eddies having a size significantly smaller than the channel width;  then the energy-containing eddies grow in size,  essentially by a classical process (see ref. \cite{davidson}, pg.347), until their size saturates at a value comparable with the width of the channel.  Recent experiments have cast doubt on the supposed details of this evolution of the energy-containing eddies, but,  as we shall see,  there seems now to be little doubt that eventually the turbulence settles down to a quasi-steady state in which the energy-containing eddies have a fixed size,  determined by the channel width,  and in which there is an inertial sub-range characterized by a Kolmogorov energy spectrum,  terminated by dissipation described by Eq.(\ref{Eq6}).  In the case of the decay of TCT when the steady heat flux is large, the initial stages are complicated,  as we outlined in our introduction,  but again there is little doubt that eventually the turbulence settles down to a state similar to that seen in the decay of grid turbulence.

As was shown first by Stalp \textit{et al} \cite{Stalp1999a},  the decay of vortex line density in the state to which the turbulence settles down is given by

\begin{equation}   L(t) = \frac{(3C)^{3/2} D}{2\pi\kappa\nu'^{1/2}}(t-t_0)^{-3/2},
\label{Eq14}
\end{equation}
where $C$ is the Kolmogorov constant,  $D$ is the (time-independent) linear size of the energy-containing eddies,  and $t_0$ is a constant. Eq.(\ref{Eq14}) is based on an assumed energy spectrum that has the Kolmogorov form with a simple cut-off for wave number less than $2\pi/D$.   Until recently,  all measurements of the effective kinematic viscosity,  $\nu'$, have been based on observations of $L(t)$ and the assumption that $D$ is exactly equal to the width of the channel in which the flow is taking place. The questionable assumptions underlying this work meant that the values of $\nu'$ were,  at best, uncertain to within a factor of perhaps two or three.  Furthermore,  since the effective size of the energy-containing eddies could depend on the precise way in which the turbulence was generated,  values of $\nu'$ from different experiments might not agree.

\begin{figure}[ht]
\begin{center}
\includegraphics[scale=0.62]{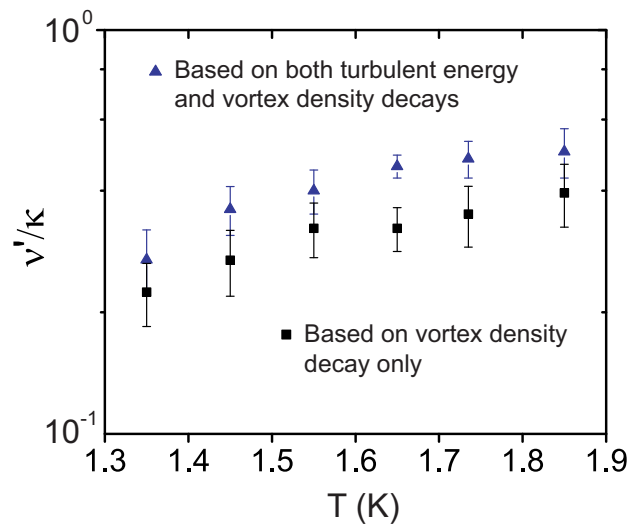}
\caption{(color online) Values of $\nu'$ for decaying TCT derived from measurements of both the decaying turbulent energy and the decaying vortex line density.  Values of $\nu'$ derived from the decay of line density alone,  based on Eq.(\ref{Eq14}) are included for comparison.}
\label{Fig11}
\end{center}
\end{figure}

This uncertainty can be circumvented if a measurement of $L(t)$ is combined with a measurement of the way in the total turbulent energy decays, since this decay of total energy yields the value of the energy flux, $\epsilon$, in Eq.(\ref{Eq6}).  The time-dependence of the total energy can be deduced from the recently developed visualization technique based on the use of He$_2^*$ excimer molecules as tracers, provided that it is assumed that the turbulence is isotropic. The first such study,  on the decay of TCT,  was reported recently \cite{Gao2016a},  and the resulting values of $\nu'$ are displayed in Fig.\ref{Fig11},  along with values of $\nu'$ derived from the same measurements of the decay of line density,  but using Eq.(\ref{Eq14}) (all these measurements relate to a channel with square cross section, 9.5 mm $\times$ 9.5 mm, and $D$ in Eq.(\ref{Eq14}) was taken to be 9.5 mm).  We see that the measurements based on the new technique are systematically slightly larger than those based on Eq.(\ref{Eq14}),  but only by a factor that is barely outside the experimental error.

\begin{figure}[ht]
\begin{center}
\includegraphics[scale=0.6]{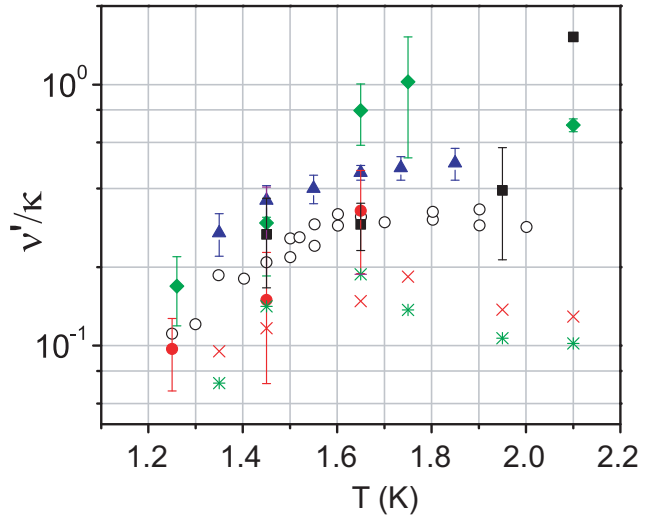}
\caption{(color online) Values of $\nu'$ for various types of decaying coupled turbulence. \textcolor{blue}{$\blacktriangle$}: ref \cite{Gao2016a};  $\circ$: ref \cite{Stalp1999a}; \textcolor{green}{$\blacklozenge$}: ref \cite{babuin2015a} decay of superflow in channel D7; \textcolor{red}{$\bullet$}: ref \cite{babuin2015a} decay of superflow in channel D10; $\blacksquare$: ref \cite{babuin2015a} decay of counterflow in channel D10; \textcolor{red}{$\times$}: ref \cite{babuin2014b} no grid; \textcolor{green}{$\ast$}: ref \cite{babuin2014b} with grid.}
\label{Fig12}
\end{center}
\end{figure}

In Fig.\ref{Fig12} we collect together the results of measurements of $\nu'$ for various types of decaying coupled quantum turbulence,  as described in the caption to the figure.  Most of these data were derived from measurements on line density only,  and for this reason are subject to some uncertainty.  There is a hint that the value of $\nu'$ may depend a little on the type of flow,  but the relatively large experimental errors make it hard to be sure.  All that we can say is that $\nu'/\kappa$ lies in the range 0.1 to 1,  its value increasing as the temperature increases from 1.3K to 2.1K.

These measurements of $\nu'$ have all been based on the decay of the quantum turbulence.  Some information about $\nu'$ has also been obtained from observations of vortex line density in the steady flow of superfluid $^4$He through a channel or through a grid \cite{babuin2014b}.  In essence,  it was tentatively assumed that the steady flow led to the generation of large eddies,  size $D$ and characteristic velocity $U$. The velocity $U$ is assumed to be proportional to the average steady flow velocity $U_0$   ($U=\zeta U_0$, where the constant $\zeta$ is a little less than unity),  and $D$ is assumed to be independent of $U_0$.  The large eddies are assumed to decay through a cascade at a rate determined by the turnover time $D/U$,  the energy being ultimately dissipated at a rate given by Eq.(\ref{Eq6}).  These assumptions lead then to a steady vortex line density given by
\begin{equation} L= \zeta^{3/2}\frac{1}{(\nu'\kappa D)^{1/2}}U_0^{3/2}.
\label{Eq15}
\end{equation}
That $L$ is proportional to $U_0^{3/2}$ is confirmed by experiment.  Eq.(\ref{Eq15}) can then be used to estimate $\nu'$, subject to reasonable guesses about the values of $\zeta$ and $D$. The results are not inconsistent with those described above,  demonstrating that the concept of an effective kinematic viscosity is applicable to dissipation in both steady and decaying turbulence;  but reliable absolute values of $\nu'$ cannot be deduced.

\section{Dissipation in quasi-classical quantum turbulence:  the relation between $\nu'_v$ and $\nu'$}

\subsection{Introduction}

We devote this section to an introductory discussion of the relation between $\nu'_v$, derived from our values of $\chi_2$, and $\nu'$.  We have already emphasized that these two kinematic viscosities relate to different physical situations,  and that they may not therefore be equal.

In the case of $\nu'_v$ we are dealing with a situation where turbulent energy in the superfluid component is confined to scales of order $\ell$ or less, in the form of a random vortex tangle,  and we assumed in our earlier discussion that there was no turbulent motion of the normal fluid. As we have seen, turbulent energy is then being dissipated by mutual friction,  at a rate that is given to a reasonable approximation by the prediction of Eq.(\ref{Eq12}). In the case of $\nu'$ there is again turbulent energy in the superfluid component on scales of order $\ell$ or less, but this is accompanied by turbulent energy in both fluids at larger scales. On sufficiently large scales there is strong coupling between the two fluids,  and viscosity in the normal fluid can be neglected.  There is then a Kolmogorov (inertial range) energy spectrum in this coupled motion,  leading to constant fluxes of energy in $k$-space in both the superfluid and normal components ($\epsilon_s$ and $\epsilon_n$).  We must discuss how this situation changes as the scale of the turbulence moves towards the scale $\ell$;  in other words how the energy spectra for the two fluids behave as the wave number approaches $2\pi/\ell$.  In connection with dissipation,  we need ultimately to answer several questions.  How,  and at what wave numbers, is turbulent energy in the normal fluid dissipated,  remembering that such dissipation can be due to both viscosity and mutual friction?  Is there significant dissipation in the superfluid component due to mutual friction at wave number smaller than $2\pi/\ell$?  And how is dissipation in the superfluid component modified,  in comparison with that for a random vortex tangle,  for wave numbers of order or greater than $2\pi/\ell$,  by any motion on those scales of the normal fluid or by the polarization of the tangle required to generate the large-scale turbulence.

\subsection{Guidance from the calculations of Bou\'e \textit{et al}}

These questions can be answered to some degree by appealing to the work of Bou\'e \textit{et al} \cite{Lvov2015a}. These authors used a two-fluid Sabra shell-model,  based on modified HVBK equations,  to calculate the energy spectra for both the superfluid and normal components.  The HVBK equations are course-grained (continuum) equations of motion for the two fluids,  and Bou\'e \textit{et al} modify them by the addition of an effective kinematic viscosity,  equal to our $\nu'$,  to the equation for the superfluid component.  Our $\nu'$ is indeed an effective kinematic viscosity in the sense that the associated dissipation,  equal to $\nu'\kappa^2 L^2$,  appears to be analogous to the classical dissipation $\nu \langle\omega^2\rangle$,  where $\langle\omega^2\rangle$ is the mean square classical vorticity. However, this analogy is misleading because our $\nu'$ is actually due,  at least in part,  to mutual friction,  so that its effect ought not to be represented by a term of the form $\nu' \nabla^2 v_s$,  as assumed by Bou\'e \textit{et al}.  Leaving aside this questionable aspect of the analysis by Bou\'e \textit{et al},  there is still the assumption that course-grained equations of motion can be used.  This assumption is probably justified in describing turbulence on scales large compared with the vortex line spacing, $\ell$,  but Bou\'e \textit{et al} use it on scales as small as the vortex line spacing.  It must fail on such small scales,  although the precise scale below which it fails noticeably is not clear. We shall return to this point later.

In spite of these shortcomings,  it is interesting to examine the conclusions to be drawn from the analysis of Bou\'e \textit{et al},  especially as they relate to the effect of the finite viscosity of the normal fluid in the range of temperatures in which we are interested (Fig.1(b) in ref \cite{Lvov2015a}). We find that, at temperatures less than about 1.5 K,  the normal fluid is brought to rest by its viscosity on a length scale significantly larger than $\ell$, but that the superfluid is then brought to rest only on significantly smaller scales, comparable with $\ell$. At first sight this is surprising,  because it might be thought that with the normal fluid at rest the superfluid would also be brought to rest by mutual friction.  There is,  however,  a simple explanation.  On scales larger than $\ell$ there is according to Bou\'e \textit{et al} a flow of energy in the turbulent superfluid to higher wave numbers in a Richardson cascade. If the normal fluid is at rest,  this cascade has associated with it two characteristic times: the turnover time for eddies centred on wave number $k$,  which is of order $\tau_t=(ku)^{-1}$,  where $u$ is the characteristic velocity in these eddies; and the time taken for the energy in these eddies to be dissipated by mutual friction,  which is of order $\tau_{\gamma}=\ell^2/\alpha\kappa$.  If $\tau_t \ll \tau_{\gamma}$,  then the mutual friction has little effect.  It is easy to show that this condition is indeed satisfied in the cases we are considering.

At temperatures above about 1.5K Bou\'e \textit{et al} show that energy is lost from both the normal component and the superfluid component only on length scales comparable with $\ell$.  It follows then that at all temperatures relevant to the present study turbulent energy is lost from the superfluid component only on length scales comparable with the vortex line spacing, $\ell$.  We emphasize that this conclusion is dependent on the questionable assumption that turbulence in the superfluid component is behaving quasi-classically on scales down to a value close to the vortex line spacing $\ell$.

\subsection{Dissipation in the superfluid component}

If we accept this assumption,  then we can conclude that, even in quasi-classical quantum turbulence of the type we are considering, energy is dissipated in the superfluid component only on length scales comparable with the vortex line spacing $\ell$,  as is the case when we have only a random vortex tangle.  It is therefore tempting to conclude that the dissipation in the superfluid component is still given by the theory of Section III.  However,  two questions must still be addressed.  The first relates to the fact that, according to Bou\'e \textit{et al},  and in contrast to the assumptions in Section III, there is motion of the normal fluid on scales of order $\ell$,  at least at the higher temperatures. But it seems reasonable to assume that on this scale the vortex line velocity given by the local induction approximation,  which is dominated by quantum effects,  is uncorrelated with the velocity field of the normal fluid,  which is essentially classical. In this case the theory of Section III still holds. The second relates to the fact that in our quasi-classical quantum turbulence the vortex lines must be moving at a velocity that includes a quasi-classical component,  corresponding to a large scale quasi-classical velocity field arising from a partial polarization of the lines.  This component is associated with the long-range, non-local, interaction of the vortex lines, and the large-scale coupling between the two fluids ensures that this component is not subject to any dissipation by mutual friction. But the fact that the argument of Section III is based on the local induction approximation ensures that this component is automatically excluded from any contribution to the dissipation (although the existence of the large-scale motion may influence the value of $c_2$).

We conclude then that the dissipation in the superfluid component in quasi-classical quantum turbulence may still given correctly by the theory of Section III,  subject,  of course, to the assumption implicit in the work of Bou\'e \textit{et al} that turbulence in the superfluid component can be regarded as quasi-classical on all scales larger than $\ell$.

\subsection{The total energy dissipation}

To obtain the total energy dissipation we must add the energy dissipated in the normal fluid.  We note that at small wave numbers,  within the inertial range, where there is complete coupling between the two fluids,  the energy fluxes in the normal and superfluid components must be given respectively by $\epsilon_n = (\rho_n/\rho)\epsilon$ and $\epsilon_s= (\rho_s/\rho)\epsilon$,  where $\epsilon$ is the total energy flux.  It follows that the effective kinematic viscosity $\nu'$,  describing the total dissipation,  is given by
\begin{equation}  \frac{\nu'}{\kappa} = \frac{\alpha c_2^2}{16\pi^2}\Big[\ln{\frac{\ell}{\xi_0}}\Big]^2 = \frac{\chi_2}{8\pi^2} \ln{\frac{\ell}{\xi_0}}
\label{Eq21}
\end{equation}
We emphasize that, as is the case with $\nu'_v$,  there is a strong dependence on the parameter $c_2$,  to which we shall return.

\subsection{Comparison with experiment}

In principle Eq.(\ref{Eq21}) serves to predict both the value of $\nu'$ and the relation between $\nu'$ and $\chi_2$ (or $\nu'_v$).   The latest experimental data displayed in Figs. \ref{Fig3} and \ref{Fig11} are, within large experimental errors,  more or less consistent with the predicted relation between $\nu'$ and $\chi_2$. However, such agreement has in practice little real significance,  because all three dissipative coefficients depend on the parameter $c_2$, the precise value of which depends on the details of the flow concerned.  Strictly speaking,  our demonstration that $c_2$ depends on these details has been established by simulations that relate only to particular flows in which the normal fluid is not turbulent, and for which the density of vortex line is small. These flows rarely correspond to reality,  especially when we are dealing with flows at high velocities that involve turbulence in both fluids and high densities of vortex line. Although the development of simulations that relate to these more general conditions has started,  we can be fairly certain that full development will many years.  In the meantime we must assume that the dependence of $c_2$ on the details of the flow is quite general.  The consequences are particularly serious for the value of $\nu'$, since the flows to which $\nu'$ is applicable are as yet the least accessible to realistic simulation.

In comparing experiment and theory relating to quasi-classical quantum turbulence we must also recognize,  as we have already emphasized,  that the theory is based on the questionable assumption made by Bou\'e \textit{et al} that the turbulence in the superfluid component behaves classically (in effect that the discrete vortex structure is unimportant) even when the length scale is comparably with $\ell$. Perhaps \textit{fully} classical behaviour may not be required,  but at least there must still be something equivalent to a Richardson cascade,  with "eddies" that have lifetimes sufficiently small that they are not damped significantly by mutual friction with a stationary normal fluid.  We guess that justification of this assumption can come only from suitable numerical simulations.  It is our plan to attempt such simulations in the near future.

\section{Summary and conclusions}

We have summarised what we know from experiment about dissipation in quantum turbulence in superfluid $^4$He above 1K, the dissipation being described by either the parameter $\chi_2$,  applicable to turbulence existing in the superfluid component only on scales comparable with the spacing between the quantized vortex lines (``random tangles") ,  or the effective kinematic viscosity $\nu'$ applicable to quasi-classical quantum turbulence,  such as that generated by flow through a grid. Theoretical predictions for these two parameters are discussed,  and it is argued that both depend on the dissipative effects of mutual friction,  which are in turn dependent on the dimensionless parameter $c_2$ that relates the mean square curvature of the vortices to their mean square separation. This parameter can in principle be obtained from simulations,  but it is argued that simulations that are sufficiently realistic are for the most part not yet practicable.  To this extent our understanding of dissipation in quantum turbulence in $^4$He above 1K remains incomplete.

We have also drawn attention to the need to investigate more carefully than hitherto the extent to which turbulence in the superfluid component can be treated classically on length scales larger than,  but comparable with, the spacing between the vortex lines.

\vspace{8mm}

\begin{acknowledgments}
J.G. and W.G. acknowledge the support from the National Science Foundation under Grant No. DMR-1507386. The experimental part of the work was conducted at the National High Magnetic Field Laboratory, which is supported by NSF Grant No. DMR-1644779 and the state of Florida.  M.T. was supported by JSPS KAKENHI Grant No. 17K05548 and MEXT KAKENHI ``Fluctuation and Structure" Grant No.16H00807. S.Y. was supported by Grant-in-Aid for JSPS Fellow Grant No. JP16J10973.

\end{acknowledgments}

\bibliographystyle{apsrev4-1}
\bibliography{qtbib}

\end{document}